\documentclass[aps,prl,twocolumn,superscriptaddress,longbibliography]{revtex4-2}

\usepackage{amsmath,amssymb,amsfonts,bm}
\usepackage{graphicx}
\usepackage[colorlinks,linkcolor=blue,citecolor=blue]{hyperref}

\newcommand{\eq}{\begin{equation}}
\newcommand{\eqx}{\end{equation}}
\newcommand{\eqn}{\begin{eqnarray}}
\newcommand{\eqnx}{\end{eqnarray}}
\newcommand{\f}[2]{\frac{#1}{#2}}

\newcommand{\cor}[1]{\left\langle{#1}\right\rangle}

\renewcommand{\SS}{{\mathcal{S}}}
\newcommand{\qq}{\quad\quad}

\begin{document}

\title{Collective Organization of String Fragmentation}

\author{Romuald A. Janik}
\affiliation{Institute of Theoretical Physics and Mark Kac Center for Complex Systems Research,\\
Jagiellonian University, Kraków, Poland}

\author{Maciej A. Nowak}
\affiliation{Institute of Theoretical Physics and Mark Kac Center for Complex Systems Research,\\
Jagiellonian University, Kraków, Poland}

\author{Marek M. Rams}
\affiliation{Institute of Theoretical Physics and Mark Kac Center for Complex Systems Research,\\
Jagiellonian University, Kraków, Poland}

\author{Ismail Zahed}
\affiliation{Center for Nuclear Theory, Department of Physics and Astronomy,
Stony Brook University, Stony Brook, New York 11794, USA}


\begin{abstract}
How coherent confining strings reorganize into hadronic matter is a central open problem in nonequilibrium gauge dynamics.
Using real-time tensor-network simulations of the massive Schwinger model, we identify a collective organization underlying string fragmentation. For intermediate masses, the dynamics simplifies and more than $99\%$ of the temporal evolution of the electric-field probability distribution is governed by a single collective mode whose decay tracks the survival probability of the parent string. The extracted survival law exhibits the same Gaussian dependence implied by the Lund world-sheet area law, providing microscopic support for its characteristic area law dependence without assuming the Lund fragmentation mechanism.
The same mode controls the transition from the coherent string to the produced mesons, explains the universal non-thermal effective fluid equation of state, and essentially controls the growth of bipartite entanglement through the causal propagation of massive pairs. These results identify collective fragmentation as an organizing principle of coherent string breaking and provide a microscopic perspective on the dynamical origin of a central ingredient of Lund hadronization.
\end{abstract}

\maketitle

\paragraph{Introduction.}
The conversion of coherent color flux into hadrons is one of the most fundamental nonequilibrium processes in quantum field theory.  It is the microscopic origin of hadronization in high-energy collisions and underlies the formation of jets, where confining strings created between luminally separating color charges fragment into color-singlet hadrons. The phenomenological description of this process is remarkably successful, but the dynamical mechanism by which a coherent gauge field reorganizes into hadronic matter remains poorly understood.

The standard description is provided by the Lund fragmentation model, in which string breaking is governed by a local probability per unit invariant world-sheet area~\cite{Andersson:1983ia,Andersson:1998tv}. Related two-dimensional studies have long provided a useful laboratory for confinement, string dynamics, inclusive production, and fragmentation~\cite{tHooft:1974pnl,Casher:1974vf,Banks:1976ia,Bardeen:1975gx,Bars:1976nk,Einhorn:1976uz,Einhorn:1976ev,Shuryak:2021hng}. However, the Lund area law is phenomenological rather than a consequence derived directly from the underlying real-time quantum dynamics. Establishing how such a law can emerge microscopically remains an outstanding problem.

The Schwinger model offers a particularly clean setting in which to address this question. From the original solution of massless QED$_2$~\cite{Schwinger:1962tp}, its massive extension, bosonized description, and particle spectrum have been extensively studied~\cite{Coleman:1974bu,Coleman:1976uz,Bergknoff:1976xr,Smilga:1992hx}. In Hamiltonian lattice form~\cite{Banks:1975gq,Kogut:1974ag,Susskind:1976jm,Dempsey:2022nys}, the theory contains precisely the ingredients relevant here: confinement, Schwinger pair production, screening, string breaking, and meson formation.

These features have made QED$_2$ a benchmark for modern nonperturbative real-time methods. Matrix-product-state and tensor-network formulations~\cite{Banuls:2013jaa,PhysRevLett.113.091601,Haegeman:2011zz} have opened direct access to gauge-theory dynamics in Minkowski space, including string breaking and pair production~\cite{Pichler:2015yqa,Buyens:2016ecr,Martinez:2016yna,Klco:2018kyo,Zache:2018cqq,Rigobello:2021fxw}. In parallel, digital and variational quantum-simulation approaches have developed rapidly~\cite{Peruzzo:2014dbu,Kandala:2017vok,Kreshchuk:2020aiq,Qian:2021jxp,Farrell:2023fgd,Zache:2023cfj,Farrell:2024fit,Gustafson:2024bww}, and recent works extended the real-time study of the Schwinger model to collisions, open-system dynamics, jet production, and hadronization~\cite{Lee:2023urk,Belyansky:2023rgh,Florio:2023dke,Florio:2024aix,Lin:2024eiz,Barata:2023jgd}.

A closely related development is the use of entanglement as a dynamical probe of pair production, confinement, and expanding quantum matter~\cite{Berges:2017hne,Berges:2017zws,Florio:2021xvj,Rigobello:2021fxw,Florio:2023mzk,Grieninger:2023ufa,Grieninger:2023pyb,Grieninger:2023ehb}. More recent studies have pushed these connections to partonic observables, real-time hadron dynamics, and quantum-computational descriptions of QED$_2$~\cite{Grieninger:2024cdl,Florio:2025hoc,Shao:2025obi,Shao:2025ygy,Grieninger:2025rdi,Artiaco:2025qqq,Galvez-Viruet:2025rmy,Galvez-Viruet:2025ket}. Together, these developments make the massive Schwinger model a controlled microscopic laboratory for asking how fragmentation, collective dynamics, and entanglement are organized during hadronization.

\begin{figure}[t]
\centering
\includegraphics[width=0.85\columnwidth]{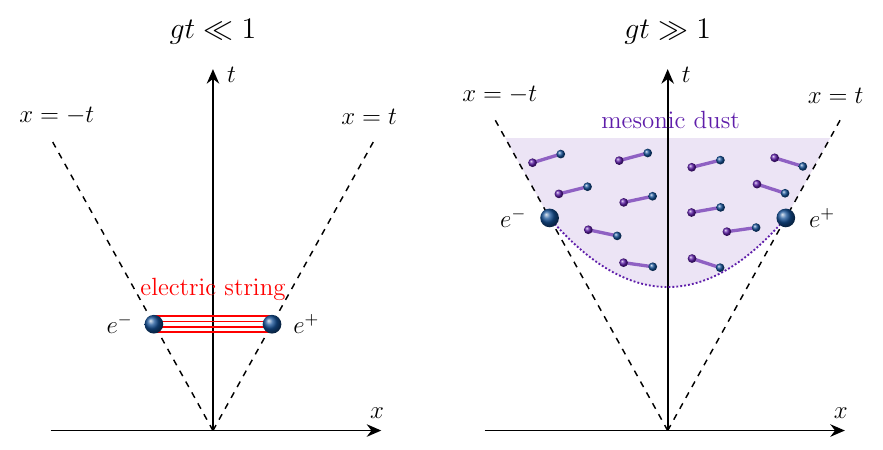}
\caption{Space-time picture of string fragmentation in the massive Schwinger model. At $gt\ll1$, receding external charges on $x=\pm t$ are joined by a coherent electric string; at $gt\gg1$, the string has fragmented into an approximately boost-invariant, weakly interacting mesonic dust.}
\label{fig:setup}
\end{figure}

\begin{figure*}[t]
\includegraphics[width=\textwidth]{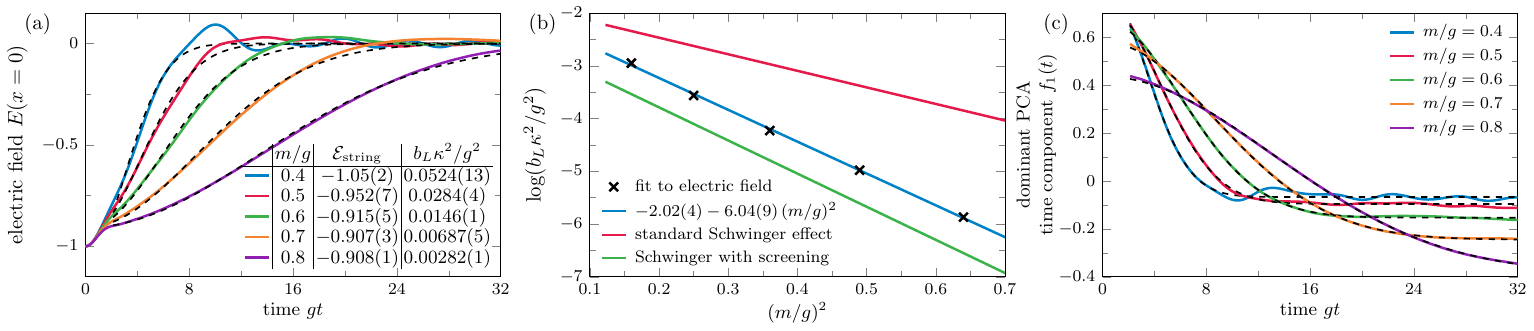}
\caption{In (a), we show normalized coherent-string survival factor, ${\cal S}(t)$, from the midpoint electric field. Tensor-network data are compared with the Gaussian Lund-area-law fit, Eq.~(\ref{eq:gaussian}). In (b), mass dependence of the string breaking coefficient is compared to Schwinger effect estimate with and without screening. In (c), we show dominant principal-component amplitude, Eq.~\eqref{eq:onemode}, of (multidimensional-) midpoint electric field probability distributions. The independent Gaussian fit reproduces the string survival probability.}
\label{fig:survival}
\end{figure*}

The physical setup is illustrated in Fig.~\ref{fig:setup}.  Initially, a coherent electric string is stretched between two external charges that move along lightlike trajectories $x=\pm t$.  As the string fragments through repeated Schwinger pair production, its energy is converted into an expanding ensemble of color-singlet mesons. At late times these mesons populate the interior of the proper-time hyperbola bounded by the receding charges, forming a neutral, almost pressureless mesonic ``dust''.

In our previous work~\cite{Janik:2025bbz}, real-time tensor-network simulations showed that for intermediate masses the dynamics simplifies strongly: expanding strings evolve into an approximately free-streaming, boost-invariant mesonic medium with a quasi-universal non-thermal effective equation of state, while the evolution is simultaneously tied to the growth of entanglement.  Those results established the macroscopic organization of the process but did not identify its microscopic origin.

In this article, we identify the missing organization. At the level of the electric-field probability distribution, coherent string fragmentation is governed by a single collective mode. Rather than exploring many independent directions in observable probability space, fragmentation dynamics follows an effectively one-dimensional trajectory. We show that this same mode tracks the survival probability of the parent string, reproduces the Gaussian time-dependence characteristic of the Lund world-sheet area law, controls the transfer from the coherent string to the produced mesons, and organizes both the effective energy-momentum tensor and the growth of bipartite entanglement.

\begin{figure*}[t]
\includegraphics[width=\textwidth]{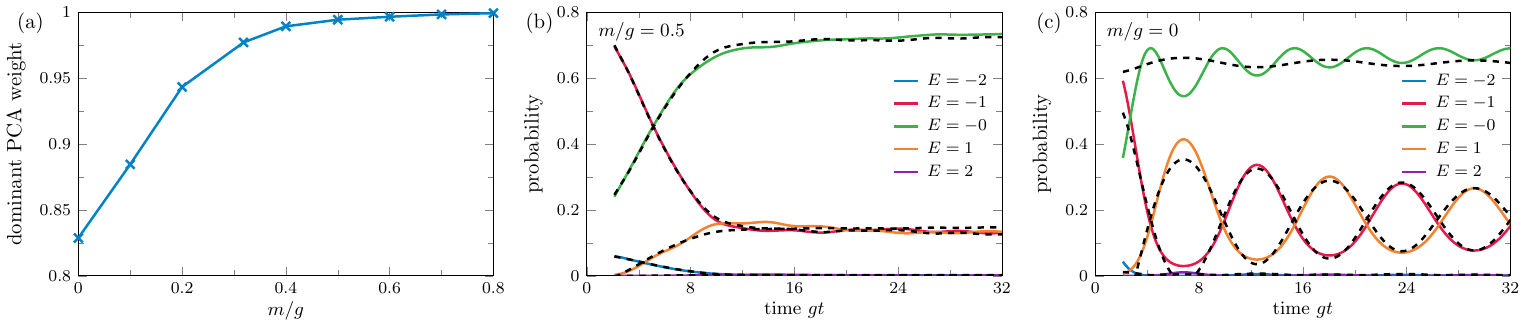}
\caption{In (a), we show the fraction of temporal variance of electric field probabilities explained by the leading PCA component. In (b), we show full tensor-network probability distribution (solid) and leading-principal-component reconstruction (dashed) for $m/g=0.5$. Their near agreement shows that one collective mode captures the observable probability redistribution.
In (c), analogous plot for $m/g=0$ clearly requires higher dimensional evolution.}
\label{fig:pca1}
\end{figure*}

\paragraph{Collective string survival mode.}
We study the real-time fragmentation of an electric string stretched between two external charges receding along light-like trajectories in the massive Schwinger model, see Fig.~\ref{fig:setup}, as is briefly reviewed in the Supplementary Material. The tensor-network formulation and its numerical implementation are described in~\cite{Janik:2025bbz}.

As the quantum mechanical wavefunction obtained from real-time tensor network evolution contains information on all possible break-up patterns of the electric string, it is interesting to study the dynamics of the electric field components of the wavefunction. Here we concentrate on the mid-point, $x=0$, between the receding charges. We will focus first on a simple probabilistic interpretation of the expectation values, and later show how a more detailed analysis of the wavefunction in the space of electric field values confirms this interpretation.

As the electric field at a given lattice site is discrete (and integer in our units with $g=1$), with an electric string between the external charges having $E=-1$ (to a first approximation, we will show later that this is a bit more involved), we may interpret the expectation value of the electric field at time $t$ in terms of a probability of the string still being present.
The simplest probabilistic model, which we will justify using PCA analysis later, is
\eq
\label{e.Sdef}
\cor{E(t)} = {\cal E}_{\rm string} \cdot \SS(t),
\eqx
where ${\cal E}_{\rm string}$ is the electric field of the string state, while $\SS(t)$ is the probability of a string being still present at $x=0$ at time $t$, i.e., a string survival probability.

Note that such a simple probabilistic interpretation is not automatic and \emph{cannot} hold for the massless Schwinger model and for small masses. Indeed, for $m=0$, $\cor{E(t)} = -J_0\left(g t/\sqrt{\pi}\right)$, with $J_0$ being a Bessel function, and then $\SS(t)$ defined by Eq.~\eqref{e.Sdef} would not be between 0 and 1.

In contrast, for the intermediate mass range, the numerical evolution throughout the fragmentation regime, $t \gtrsim 2$, is accurately described, see Fig.~\ref{fig:survival}(a), by
\begin{equation}
\SS(t)
=
\exp\!\left(-b_L\kappa^2 t^2\right),
\label{eq:gaussian}
\end{equation}
where $\kappa$ is the string tension, with ${\cal E}_{\rm string} \approx -0.9$. The initial time interval $t<2$ involves different behaviors and should be related to the formation of the string, which we do not discuss in this paper.

The Gaussian decay in Eq.~\eqref{eq:gaussian} immediately recalls the phenomenological Lund fragmentation model~\cite{Andersson:1983ia,Andersson:1998tv}, where the probability of a string splitting at proper time $\tau$ is given by
\eq
H(\Gamma)\, d\Gamma= C\, \Gamma^{a_L} e^{-b_L\, \Gamma} d\Gamma \equiv h(\tau)\, d\tau
\label{e.Hh}
\eqx
with $\Gamma = \kappa^2 \tau^2$ being related to the (momentum-)area of the wedge of the string. The string survival probability can then be estimated (for $a_L=0$) by
\eq
\SS(t) = 1 - \int_0^{t} h(s)\, ds = e^{-b_L\kappa^2 \, t^2},
\label{e.survival}
\eqx
The tensor-network simulations therefore reproduce microscopically the Gaussian dependence characteristic of the Lund survival law. This agreement concerns the area-law functional dependence and should not be interpreted as a derivation of the full Lund fragmentation model. Rather, a survival law introduced phenomenologically in Lund fragmentation emerges here from the real-time quantum dynamics, independently. Although the Lund model does not predict the numerical value of the coefficient $b_L$, it correctly captures the dynamical form of the coherent-string decay.

The coefficient itself exhibits a simple Gaussian dependence on the fermion mass,
\begin{equation}
\ln (w/g^2) = \ln (b_L\kappa^2 / g^2) = \!-2.02(4) - 6.04(9) (m/g)^2,
\label{eq:massfit}
\end{equation}
throughout the mass range investigated, see Fig.~\ref{fig:survival}(b). It is important to note that the decrease with the mass is much stronger than the conventional Schwinger production formula in an external field
\eq
\label{e.schwingerext}
w_{Schwinger} \sim \f{g |E_{\rm ext}|}{2\pi} e^{-\f{\pi m^2}{g |E_{\rm ext}|}}.
\eqx
Indeed, the exponent approximately matches a dynamical version of the Schwinger effect, where the electric field gets screened by the produced fermions~\cite{Kleban:2011cs}, leading to the substitution
$g|E_{\rm ext}| \to g|E_{\rm eff}| = g\left(|E| - \f{g}{2} \right)$
in Eq.~\eqref{e.schwingerext}. Both variants can be reinterpreted in terms of an Unruh temperature $T_U=\frac{g|E_{\rm ext}|}{2\pi m}$ or $T_U=\frac{g|E_{\rm eff}|}{2\pi m}$, as in~\cite{Gomberoff:2005je}. This reference automatically recovers the apparent mismatch factor of 2, see~\cite{Volovik:2022cqk}, in $\exp(-m/2T_U)$.

The survival factor is extracted solely from the \emph{expectation value} of the midpoint electric field. A more stringent test is whether the same collective mode also governs the evolution of the complete \emph{probability distribution}. We now demonstrate that this is indeed the case for the range of intermediate masses under consideration but fails for small masses and especially for $m=0$, in accordance with the breakdown of the probabilistic interpretation of Eq.~\eqref{e.Sdef} in that case.

\begin{figure*}[t]
\centering
\includegraphics[width=\textwidth]{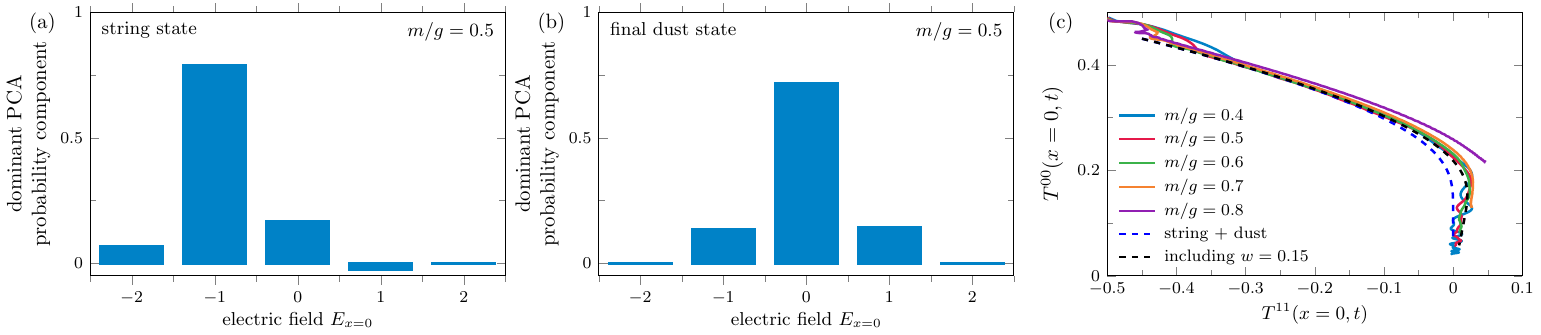}
\caption{Midpoint electric-field probability distributions at $m/g=0.5$ for, in (a), the coherent parent-string state~\cite{footnote_nonpositive} and, in (b), the fragmented mesonic (``dust'') state. Fragmentation transfers probability from the string sector to screened configurations along a single dominant direction in probability space. In (c), we show effective equation of state in the coarse-grained fragmentation model. The survival probability controls the transfer of energy from the confining string, $p\simeq-\epsilon$, to an expanding mesonic medium with a small positive longitudinal pressure.}
\label{fig:probabilities}
\end{figure*}

\paragraph{Collective fragmentation mode.}
To quantify this collective evolution, we analyze the probability distribution $p_{E}(t)$ of the midpoint electric field $E$ by decomposing the many-body state as
\begin{equation}
|\Psi(t)\rangle
=
\sum_{E}
a_{E}(t)
|\Psi_{E}(t)\rangle , \qq p_{E}(t) = |a_{E}(t)|^2.
\label{eq:state}
\end{equation}
We decompose the probability vector $\mathbf p(t)=\{p_{E}(t)\}$ into principal components~\cite{JolliffeCadima2016},
\begin{equation}
p_{E}(t)
=
\overline p_{E}
+
\sum_n
f_n(t)\,
p^{(n)}_{E},
\label{eq:pca}
\end{equation}
where $\overline p_{E}$ is the time-averaged distribution, $p^{(n)}_{E}$ the principal directions in the probability space, and $f_n(t)$ their amplitudes.

Throughout the fragmentation process, for $m/g\geq 0.5$, the leading principal component alone accounts for more than $99\%$ of the temporal variance, see Fig.~\ref{fig:survival}(a),
\begin{equation}
p_{E}(t)
\simeq
\overline p_{E}
+
f_1(t)\,
p^{(1)}_{E},
\label{eq:onemode}
\end{equation}
showing that the measured dynamics follows an effectively one-dimensional trajectory in probability space despite the exponentially large Hilbert space. This reduction characterizes the observable probability distribution, not the full many-body wave function. Its physical content is strengthened by the fact that the same mode is independently selected by the parent-string survival probability and that the reduction fails toward the massless limit.

Figure~\ref{fig:pca1}(b) compares this reconstruction with the full tensor-network evolution. In contrast, for $m/g=0$ the variance explained by the leading component is $83\%$, and indeed there is a significant additional missing component clearly seen in Fig.~\ref{fig:pca1}(c).

The dominant PCA amplitude $f_1(t)$ can be fitted in the form
\eq
f_1(t) = \alpha_0 \, e^{-b_L\kappa^2 t^2} - \alpha_1,
\eqx
where we fixed $b_L\kappa^2$ at the value obtained in Fig.~\ref{fig:survival}(a), still obtaining an excellent fit.
This leads to the following two component model for probabilities
\eq
p_{E}(t) = \SS(t)\, p_{E}^{\rm string} + (1-\SS(t))\,  p_{E}^{\rm dust},
\eqx
with $p_{E}^{\rm string}$ and $p_{E}^{\rm dust}$ given by
\eq
p_{E}^{\rm string} = \overline p_{E} + (\alpha_0-\alpha_1) p^{(1)}_{E},
\qq
p_{E}^{\rm dust} = \overline p_{E} + (0-\alpha_1) p^{(1)}_{E}.
\eqx
Evaluating $\sum_{E} E p_{E}^{\rm string}$, we recover ${\cal E}_{\rm string} = -0.955$ for $m/g=0.5$, quite close to the value obtained in the fit of the expectation value of the electric field.

Figures~\ref{fig:probabilities}(a) and (b) illustrate how this collective mode reorganizes the probability distribution. Initially, the probability is concentrated in the coherent-string sector and is subsequently transferred to screened mesonic sectors.

\paragraph{Effective non-thermal fluid equation of state and universality.}
As a first consequence of the collective mode, the resulting two-component picture of the electric-field probability distribution allows us to model the evolution of the energy-momentum tensor and quantitatively explain a surprising feature observed in~\cite{Janik:2025bbz}--a \emph{universal} equation of state describing the transition from a string to an almost free-streaming emergent mesonic medium.

As the coherent electric string decays, its energy is continuously transferred into longitudinally propagating color-singlet mesons with breakup probability $h(\tau)$, Eqs.~\eqref{eq:gaussian}--\eqref{e.survival}.

The coherent string contributes $T^{00}_{\rm string}=\kappa_{\rm eff} \SS(\tau)$ and $T^{11}_{\rm string}=-\kappa_{\rm eff} \SS(\tau)$, where we allow for the renormalization of the effective string tension of the string state, which we denote by $\kappa_{\rm eff}$. The produced mesons subsequently undergo boost-invariant longitudinal expansion. For a nearly pressureless equation of state $p=w\epsilon$ with $0\le w\ll1$, energy conservation for mesons produced at $s$ gives us $\epsilon(\tau|s) = \kappa_{\rm eff} \left(\frac{s}{\tau}\right)^{1+w}$. The pressureless mesonic dust observed in the tensor-network simulations corresponds to the leading limit $w=0$. The total energy-momentum tensor is, therefore,
\begin{align}
T^{00}(\tau)
&=
\kappa_{\rm eff} S(\tau)
+
\kappa_{\rm eff}
\int_0^\tau ds\,
h(s)
\left(\frac{s}{\tau}\right)^{1+w},
\nonumber\\
T^{11}(\tau)
&=
-\kappa_{\rm eff} S(\tau)
+
w\kappa_{\rm eff}
\int_0^\tau ds\,
h(s)
\left(\frac{s}{\tau}\right)^{1+w}.
\label{eq:EMTmedium}
\end{align}

Equation~(\ref{eq:EMTmedium}) interpolates between the coherent confining string, $p=-\epsilon$, and the emergent mesonic medium. In the dominant pressureless limit $T^{\mu\nu}=\epsilon\,u^\mu u^\nu$, while a small positive longitudinal pressure, $w>0$, parametrizes the weak residual meson interactions observed in the tensor-network simulations. Fig.~\ref{fig:probabilities}(c) shows this evolution with $\kappa_{\rm eff}=0.45$ and a small $w = 0.15$.
\begin{figure*}[t]
\centering
\includegraphics[width=\textwidth]{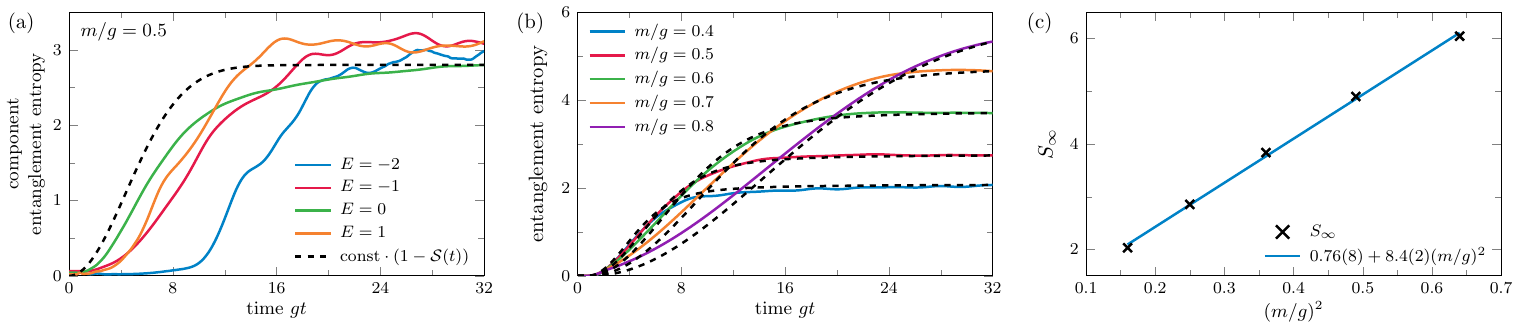}
\caption{
In (a), entanglement entropy (averaged over two neighboring sites at $x=0$) of the individual electric field components of the wave function with the mesonic dust probability for comparison ($m/g=0.5$).
In (b), bipartite entropy $S_{\rm EE}(t)$ at in the middle of the chain (solid), and the causal prediction of Eq.~(\ref{eq:entropy}) (dashed), fitting only $S_{\infty}$.
In (c), we show linear dependence between $S_{\infty}$ and $(m/g)^2$. }
\label{fig:entropy}
\end{figure*}

It is now easy to see how to recover universality. The equation of state is a parametric plot $T^{00}(\tau)$ versus $T^{11}(\tau)$. We may rescale $\tau$ by $b_L \kappa^2$, which is the only place where the dependence on the fermion mass $m/g$ enters. Therefore, the energy as a function of pressure becomes independent of the mass, as long as the assumption of a two-component system holds and the mesonic interaction is parameterized by the dimensionless parameter $w$.

\paragraph{Entanglement entropy of the string and mesonic dust.}
In order to address these questions, we project the tensor network wavefunction into components with given values of electric field at the midpoint and compute their individual entanglement entropy, see Fig.~\ref{fig:entropy}(a).

Note that we cannot \emph{directly} infer from these data the string and mesonic dust entanglement entropies, as parts of the $E=-1$ component may appear either as the genuine string state or as fluctuations of the $E=0$ mesonic state and vice versa.
However, looking at all the curves in combination, we may, nevertheless, draw some important qualitative conclusions. We focus on $t<10$, where the presence of the string should still have an impact.

First, up to $t \sim 6$, the entanglement entropies of the $E=\pm 1$ components essentially coincide. This strongly indicates that their origin arises as fluctuations around the mesonic dust state (which has symmetric $E=\pm 1$ components, see Fig.~\ref{fig:probabilities}(b)). Any significant contribution from the string should strongly modify $E=-1$ but have negligible impact on $E=1$, see Fig.~\ref{fig:probabilities}(a), thus breaking the symmetry. Second, the lack of entanglement in the string is also supported by the very flat $E=-2$ curve.

Third, the fact that the $E=0$ entanglement entropy rises approximately like $1-\SS(t)$, is also consistent with this interpretation, as it suggests that this entropy arises only from the string breakup and not from an $E=0$ fluctuation of the string which would be predominant at early times~\footnote{The string would just have a trivial, constant entropy arising from $-\sum_{E} p_{E}^{\rm string} \log_2 p_{E}^{\rm string}$ which is essentially equal to the ground state entropy.}.

\paragraph{Entanglement transfer from causal fragmentation.}
We can implement the qualitative observations above by a simple model of the entanglement entropy production, which is entirely due to the breakup of the string, dictated by the probability in Eqs.~\eqref{eq:gaussian}--\eqref{e.survival}.
A key ingredient is that the produced fermion pair contributes to the bipartite entropy only after its constituents have propagated far enough to become causally separated by the entanglement cut. Their separation at time $t$ is
\begin{equation}
L(t,t')
=
2
\int_{t'}^{t}
d\tau\,
|v(\tau;t')|,
\label{eq:length}
\end{equation}
where
\begin{equation}
v(\tau;t')
=
\frac{p(\tau;t')}
{\sqrt{m^2+p^2(\tau;t')}} ,
\qquad
p(\tau;t')
=
g
\int_{t'}^{\tau}
ds\,E(s).
\label{eq:velocity}
\end{equation}
Parameterizing the entanglement entropy formation length by $\ell_c\sim 1/\sqrt{g|E|}$~\footnote{In the ballistic regime $p\sim g|E|t$, reaching the field scale $p\sim\sqrt{g|E|}$ requires $t\sim\ell_c\sim1/\sqrt{g|E|}$.}, this leads to the model for the coarse-grained entanglement entropy
\begin{equation}
S_{\rm EE}(t)
=
S_\infty
\int_0^t
dt'
\,h(t')
F\!\left(
\frac{L(t,t')}{\ell_c}
\right),
\label{eq:entropy}
\end{equation}
where $F(0)=0$ and $F(\infty)=1$. For simplicity, we pick $F(x)=\Theta(x-1)$. The normalization $S_\infty$ characterizes the entanglement carried by a single fragmentation event.

Fig.~\ref{fig:entropy}(b) compares the resulting prediction with the tensor-network calculations for several fermion masses. The model reproduces both the delayed onset and the systematic slowing of entropy growth with increasing fermion mass. The results are only weakly sensitive to the precise choice of $\ell_c$, and we fix $\ell_c=1$ and $b_L \kappa^2$ from  Fig.~\ref{fig:survival}(a), while fitting for $S_\infty$ in Fig.~\ref{fig:entropy}(b) (for smaller masses, fitting just recovers the asymptotic value).

Interestingly, the asymptotic entropy exhibits the empirical scaling
\begin{equation}
    S_\infty\simeq 0.76(8)+8.4(2)(m/g)^2
\end{equation}
in bits, see Fig.~\ref{fig:entropy}(c). Combined with the independently
extracted fragmentation rate~\eqref{eq:massfit}, this yields
$S_\infty\simeq{\rm const}-0.96(3)\log_2(w/g^2)$.
Consequently, the effective Schmidt dimension
$N_{\rm eff}=2^{S_\infty}$ scales as
\begin{equation}
    N_{\rm eff}\sim(g\tau_{\rm br})^{1.93(5)}
    \sim\left(A_{\rm L}/\ell_s^2\right)^{0.96(3)},
\end{equation}
where $\tau_{\rm br}\sim w^{-1/2}$,
$A_{\rm L}\sim\tau_{\rm br}^2$ is the characteristic
Lund fragmentation area, and $\ell_s\sim\kappa^{-1/2}$
is the string length scale.
Remarkably, the exponent is consistent with unity.
In the Lund picture, random breakups partition the
world sheet into patches associated with mesonic
singlets, with characteristic areas set by their
invariant masses. This suggests that $N_{\rm eff}$
counts the accessible singlet fragmentation
configurations, approximately one per elementary
Lund area bit. Their near-maximal mixing would give
$S_\infty\sim\log_2(A_{\rm L}/\ell_s^2)$,
linking the saturated entanglement directly to
the stochastic geometry of string fragmentation.

\paragraph{Conclusions.}
We have shown that coherent string fragmentation in the massive
Schwinger model is governed by a single collective mode. This mode
controls the decay of the parent electric string, the transfer of
probability into screened mesonic sectors, the growth of bipartite
entanglement, and the effective energy-momentum tensor throughout the fragmentation process from the string to the emergent, weakly interacting mesonic medium.
The resulting description explains the universality, i.e., the independence of the fermion mass,
and the precise shape of the equation of state of the effective non-thermal fluid observed in~\cite{Janik:2025bbz}.

The agreement between the independently extracted string-survival law
and the dominant principal component shows that the macroscopic
observables found previously are manifestations of a common collective
fragmentation dynamics at the level of the electric-field probability
distribution. The resulting Gaussian survival probability mirrors the
Lund world-sheet area law. This does not constitute a derivation of the
full Lund model; rather, it shows that its characteristic survival-law
dependence can emerge independently from microscopic real-time gauge
dynamics.

It is important to note, that the above collective simplification of the dynamics occurs at intermediate masses (possibly also for large masses which we cannot simulate due to resulting timescales), or equivalently intermediate coupling as only the ratio $m/g$ is relevant.
This range of parameters is far from weak coupling both in the bosonized and in the original fermionic formulation of the massive Schwinger model.
It will be interesting to determine whether an analogous low-dimensional collective fragmentation manifold persists in higher-dimensional confining gauge theories, and underlies realistic string hadronization.

\begin{acknowledgments}
I.Z. thanks the Mark Kac Center for Complex Systems
Research for hospitality during the completion of this work.
R.A.J., M.A.N.,
M.M.R. and I.Z. are supported by the Priority Research Area DigiWorld under the
Strategic Programme Excellence Initiative at the Jagiellonian
University. M.M.R. acknowledges support from the National Science
Centre (NCN), Poland, Grant No.~2020/38/E/ST3/00150. R.A.J.
acknowledges support from the National Science Centre (NCN), Poland,
Grant No.~2025/57/B/ST2/02102. I.Z. is supported in part by the U.S. Department of Energy, Office of
Science, under Contract No.~DE-FG02-88ER40388.
\end{acknowledgments}

\bibliography{main.bib}

@footnote{footnote_nonpositive,
note= {The minuscule negative probability for $E=1$ comes from the fact that PCA is agnostic to positivity. It is indeed appealing that the violation of positivity is so small.}
}

@article{Schwinger:1962tp,
  author = {Schwinger, Julian},
  title = {Gauge Invariance and Mass. II},
  journal = {Phys. Rev.},
  volume = {128},
  pages = {2425--2429},
  year = {1962},
  doi = {10.1103/PhysRev.128.2425}
}

@article{Kogut:1974ag,
  author = {Kogut, John B. and Susskind, Leonard},
  title = {Hamiltonian Formulation of Wilson's Lattice Gauge Theories},
  journal = {Phys. Rev. D},
  volume = {11},
  pages = {395--408},
  year = {1975},
  doi = {10.1103/PhysRevD.11.395}
}

@article{Andersson:1983ia,
  author = {Andersson, Bo and Gustafson, G. and Ingelman, G. and Sjostrand, T.},
  title = {Parton Fragmentation and String Dynamics},
  journal = {Phys. Rept.},
  volume = {97},
  pages = {31--145},
  year = {1983},
  doi = {10.1016/0370-1573(83)90080-7}
}

@book{Andersson:1998tv,
  author = {Andersson, Bo},
  title = {The Lund Model},
  publisher = {Cambridge University Press},
  address = {Cambridge},
  year = {1998}
}

@article{Pichler:2015yqa,
  author = {Pichler, T. and Dalmonte, M. and Rico, E. and Zoller, P. and Montangero, S.},
  title = {Real-time Dynamics in U(1) Lattice Gauge Theories with Tensor Networks},
  journal = {Phys. Rev. X},
  volume = {6},
  pages = {011023},
  year = {2016},
  doi = {10.1103/PhysRevX.6.011023},
}

@article{Martinez:2016yna,
  author = {Martinez, E. A. and others},
  title = {Real-time dynamics of lattice gauge theories with a few-qubit quantum computer},
  journal = {Nature},
  volume = {534},
  pages = {516--519},
  year = {2016},
  doi = {10.1038/nature18318},
}

@article{Kasper:2016mzj,
  author = {Kasper, V. and Hebenstreit, F. and Jendrzejewski, F. and Oberthaler, M. K. and Berges, J.},
  title = {Implementing quantum electrodynamics with ultracold atomic systems},
  journal = {New J. Phys.},
  volume = {19},
  pages = {023030},
  year = {2017},
  doi = {10.1088/1367-2630/aa54e0}
}

@article{Buyens:2016ecr,
  author = {Buyens, Boye and Haegeman, Jutho and Hebenstreit, Florian and Verstraete, Frank and Van Acoleyen, Karel},
  title = {Real-time simulation of the Schwinger effect with Matrix Product States},
  journal = {Phys. Rev. D},
  volume = {96},
  pages = {114501},
  year = {2017},
  doi = {10.1103/PhysRevD.96.114501},
}

@article{Klco:2018kyo,
  author = {Klco, N. and Dumitrescu, E. F. and McCaskey, A. J. and Morris, T. D. and Pooser, R. C. and Sanz, M. and Solano, E. and Lougovski, P. and Savage, M. J.},
  title = {Quantum-Classical Computation of Schwinger Model Dynamics using Quantum Computers},
  journal = {Phys. Rev. A},
  volume = {98},
  pages = {032331},
  year = {2018},
  doi = {10.1103/PhysRevA.98.032331},
}

@article{Zohar:2015hwa,
  author = {Zohar, Erez and Cirac, J. Ignacio and Reznik, Benni},
  title = {Quantum Simulations of Lattice Gauge Theories using Ultracold Atoms in Optical Lattices},
  journal = {Rept. Prog. Phys.},
  volume = {79},
  pages = {014401},
  year = {2016},
  doi = {10.1088/0034-4885/79/1/014401}
}

@article{Bauer:2022hpo,
  author = {Bauer, Christian W. and others},
  title = {Quantum Simulation for High-Energy Physics},
  journal = {PRX Quantum},
  volume = {4},
  pages = {027001},
  year = {2023},
  doi = {10.1103/PRXQuantum.4.027001},
}

@article{Bauer:2023qgm,
  author = {Bauer, Christian W. and Davoudi, Zohreh and Klco, Natalie and Savage, Martin J.},
  title = {Quantum simulation of fundamental particles and forces},
  journal = {Nature Rev. Phys.},
  volume = {5},
  pages = {420--432},
  year = {2023},
  doi = {10.1038/s42254-023-00599-8},
}

@article{Janik:2025bbz,
  author = {Janik, Romuald A. and Nowak, Maciej A. and Rams, Marek M. and Zahed, Ismail},
  title = {Universality and Emergent Effective Fluid from Jets and String Breaking in the Massive Schwinger Model Using Tensor Networks},
  journal = {Phys. Rev. Lett.},
  volume = {135},
  pages = {211903},
  year = {2025},
  doi = {10.1103/PhysRevLett.135.211903},
}

@article{Gomberoff:2005je,
  author = {Gomberoff, Andres and Henneaux, Marc and Teitelboim, Claudio},
  title = {Decay of the Cosmological Constant: Equivalence of Quantum Tunneling and Thermal Activation in Two Spacetime Dimensions},
  journal = {Phys. Rev. D},
  volume = {71},
  pages = {063509},
  year = {2005},
  doi = {10.1103/PhysRevD.71.063509},
}

@article{Volovik:2022cqk,
  author = {Volovik, G. E.},
  title = {Schwinger pair creation: analogy with polar catastrophe},
  journal = {JETP Lett.},
  volume = {116},
  pages = {577},
  year = {2022},
}

@article{Kleban:2011cs,
    author = "Kleban, Matthew and Krishnaiyengar, Karthik and Porrati, Massimo",
    title = "{Flux Discharge Cascades in Various Dimensions}",
    doi = "10.1007/JHEP11(2011)096",
    journal = "JHEP",
    volume = "2011",
    number = "11",
    pages = "096",
    year = "2011"
}

@article{Banuls:2013jaa,
    author = "Ba{\~n}uls, Mari Carmen and Cichy, Krzysztof and Cirac, J. Ignacio and Jansen, Karl",
    title = "{The mass spectrum of the Schwinger model with Matrix Product States}",
    doi = "10.1007/JHEP11(2013)158",
    journal = "JHEP",
    volume = "2013",
    number = "11",
    pages = "158",
    year = "2013",
}

@article{JolliffeCadima2016,
    author  = {Jolliffe, Ian T. and Cadima, Jorge},
    title   = {Principal component analysis: a review and recent developments},
    journal = {Phil. Trans. R. Soc. A},
    volume  = {374},
    number  = {2065},
    pages   = {20150202},
    year    = {2016},
    doi     = {10.1098/rsta.2015.0202}
}

@article{Grieninger:2024cdl,
    author = "Grieninger, Sebastian and Ikeda, Kazuki and Zahed, Ismail",
    title = "{Quasiparton distributions in massive QED2: Toward quantum computation}",
    doi = "10.1103/PhysRevD.110.076008",
    journal = "Phys. Rev. D",
    volume = "110",
    pages = "076008",
    year = "2024"
}

@article{Coleman:1976uz,
    author = "Coleman, Sidney R.",
    title = "{More About the Massive Schwinger Model}",
    doi = "10.1016/0003-4916(76)90280-3",
    journal = "Annals Phys.",
    volume = "101",
    pages = "239",
    year = "1976"
}

@article{Coleman:1974bu,
    author = "Coleman, Sidney R.",
    title = "{The Quantum Sine-Gordon Equation as the Massive Thirring Model}",
    doi = "10.1103/PhysRevD.11.2088",
    journal = "Phys. Rev. D",
    volume = "11",
    pages = "2088",
    year = "1975"
}

@article{Smilga:1992hx,
    author = "Smilga, Andrei V.",
    title = "{On the fermion condensate in Schwinger model}",
    doi = "10.1016/0370-2693(92)90209-M",
    journal = "Phys. Lett. B",
    volume = "278",
    pages = "371--376",
    year = "1992"
}

@article{Bergknoff:1976xr,
    author = "Bergknoff, Hugh",
    title = "{Physical Particles of the Massive Schwinger Model}",
    doi = "10.1016/0550-3213(77)90204-8",
    journal = "Nucl. Phys. B",
    volume = "122",
    pages = "215--236",
    year = "1977"
}

@article{Banks:1975gq,
    author = "Banks, Tom and Susskind, Leonard and Kogut, John B.",
    title = "{Strong Coupling Calculations of Lattice Gauge Theories: (1+1)-Dimensional Exercises}",
    doi = "10.1103/PhysRevD.13.1043",
    journal = "Phys. Rev. D",
    volume = "13",
    pages = "1043",
    year = "1976"
}

@article{Susskind:1976jm,
    author = "Susskind, Leonard",
    title = "{Lattice Fermions}",
    doi = "10.1103/PhysRevD.16.3031",
    journal = "Phys. Rev. D",
    volume = "16",
    pages = "3031--3039",
    year = "1977"
}

@article{Dempsey:2022nys,
    author = "Dempsey, Ross and Klebanov, Igor R. and Pufu, Silviu S. and Zan, Bernardo",
    title = "{Discrete chiral symmetry and mass shift in the lattice Hamiltonian approach to the Schwinger model}",
    doi = "10.1103/PhysRevResearch.4.043133",
    journal = "Phys. Rev. Res.",
    volume = "4",
    pages = "043133",
    year = "2022"
}

@article{PhysRevLett.113.091601,
  title = {Matrix Product States for Gauge Field Theories},
  author = {Buyens, Boye and Haegeman, Jutho and Van Acoleyen, Karel and Verschelde, Henri and Verstraete, Frank},
  journal = {Phys. Rev. Lett.},
  volume = {113},
  pages = {091601},
  year = {2014},
  doi = {10.1103/PhysRevLett.113.091601},
}

@article{Zache:2018cqq,
    author = "Zache, T. V. and Mueller, N. and Schneider, J. T. and Jendrzejewski, F. and Berges, J. and Hauke, P.",
    title = "{Dynamical Topological Transitions in the Massive Schwinger Model with a $\theta$ Term}",
    doi = "10.1103/PhysRevLett.122.050403",
    journal = "Phys. Rev. Lett.",
    volume = "122",
    pages = "050403",
    year = "2019"
}

@article{Farrell:2023fgd,
  title = {Scalable Circuits for Preparing Ground States on Digital Quantum Computers: The Schwinger Model Vacuum on 100 Qubits},
  author = {Farrell, Roland C. and Illa, Marc and Ciavarella, Anthony N. and Savage, Martin J.},
  journal = {PRX Quantum},
  volume = {5},
  pages = {020315},
  year = {2024},
  doi = {10.1103/PRXQuantum.5.020315},
}

@article{Farrell:2024fit,
  title = {Quantum simulations of hadron dynamics in the Schwinger model using 112 qubits},
  author = {Farrell, Roland C. and Illa, Marc and Ciavarella, Anthony N. and Savage, Martin J.},
  journal = {Phys. Rev. D},
  volume = {109},
  pages = {114510},
  year = {2024},
  doi = {10.1103/PhysRevD.109.114510},
}

@article{Zache:2023cfj,
    author = "Zache, Torsten V. and Gonz\'alez-Cuadra, Daniel and Zoller, Peter",
    title = "{Fermion-qudit quantum processors for simulating lattice gauge theories with matter}",
    doi = "10.22331/q-2023-10-16-1140",
    journal = "Quantum",
    volume = "7",
    pages = "1140",
    year = "2023"
}

@article{Lin:2024eiz,
    author = {Lin, Joshua and Luo, Di and Yao, Xiaojun and Shanahan, Phiala E.},
    title = {{Real-time Dynamics of the Schwinger Model as an Open Quantum System with Neural Density Operators}},
    journal = {JHEP},
    volume = "2024",
    number = "06",
    pages = {211},
    year = {2024},
    doi = {10.1007/JHEP06(2024)211}
}

@article{Florio:2023dke,
    author = "Florio, Adrien and Frenklakh, David and Ikeda, Kazuki and Kharzeev, Dmitri and Korepin, Vladimir and Shi, Shuzhe and Yu, Kwangmin",
    title = "{Real-Time Nonperturbative Dynamics of Jet Production in Schwinger Model: Quantum Entanglement and Vacuum Modification}",
    doi = "10.1103/PhysRevLett.131.021902",
    journal = "Phys. Rev. Lett.",
    volume = "131",
    pages = "021902",
    year = "2023"
}

@article{Florio:2024aix,
  title = {Quantum real-time evolution of entanglement and hadronization in jet production: Lessons from the massive Schwinger model},
  author = {Florio, Adrien and Frenklakh, David and Ikeda, Kazuki and Kharzeev, Dmitri and Korepin, Vladimir and Shi, Shuzhe and Yu, Kwangmin},
  journal = {Phys. Rev. D},
  volume = {110},
  pages = {094029},
  year = {2024},
  doi = {10.1103/PhysRevD.110.094029},
}

@article{Florio:2023mzk,
    author = "Florio, Adrien",
    title = "{Two-fermion negativity and confinement in the Schwinger model}",
    doi = "10.1103/PhysRevD.109.L071501",
    journal = "Phys. Rev. D",
    volume = "109",
    pages = "L071501",
    year = "2024"
}

@article{Grieninger:2023ufa,
    author = "Grieninger, Sebastian and Ikeda, Kazuki and Kharzeev, Dmitri E. and Zahed, Ismail",
    title = "{Entanglement in massive Schwinger model at finite temperature and density}",
    doi = "10.1103/PhysRevD.109.016023",
    journal = "Phys. Rev. D",
    volume = "109",
    pages = "016023",
    year = "2024"
}

@article{Rigobello:2021fxw,
    author = "Rigobello, Marco and Notarnicola, Simone and Magnifico, Giuseppe and Montangero, Simone",
    title = "{Entanglement generation in (1+1)D QED scattering processes}",
    doi = "10.1103/PhysRevD.104.114501",
    journal = "Phys. Rev. D",
    volume = "104",
    pages = "114501",
    year = "2021"
}

@article{Lee:2023urk,
    author = "Lee, Kyle and Mulligan, James and Ringer, Felix and Yao, Xiaojun",
    title = "{Liouvillian dynamics of the open Schwinger model: String breaking and kinetic dissipation in a thermal medium}",
    doi = "10.1103/PhysRevD.108.094518",
    journal = "Phys. Rev. D",
    volume = "108",
    pages = "094518",
    year = "2023"
}

@article{Belyansky:2023rgh,
    author = "Belyansky, Ron and Whitsitt, Seth and Mueller, Niklas and Fahimniya, Ali and Bennewitz, Elizabeth R. and Davoudi, Zohreh and Gorshkov, Alexey V.",
    title = "{High-Energy Collision of Quarks and Mesons in the Schwinger Model: From Tensor Networks to Circuit QED}",
    doi = "10.1103/PhysRevLett.132.091903",
    journal = "Phys. Rev. Lett.",
    volume = "132",
    pages = "091903",
    year = "2024"
}

@article{Kreshchuk:2020aiq,
    author = "Kreshchuk, Michael and Jia, Shaoyang and Kirby, William M. and Goldstein, Gary and Vary, James P. and Love, Peter J.",
    title = "{Simulating Hadronic Physics on NISQ devices using Basis Light-Front Quantization}",
    doi = "10.1103/PhysRevA.103.062601",
    journal = "Phys. Rev. A",
    volume = "103",
    pages = "062601",
    year = "2021"
}

@article{Qian:2021jxp,
    author = "Qian, Wenyang and Basili, Robert and Pal, Soham and Luecke, Glenn and Vary, James P.",
    title = "{Solving hadron structures using the basis light-front quantization approach on quantum computers}",
    doi = "10.1103/PhysRevResearch.4.043193",
    journal = "Phys. Rev. Res.",
    volume = "4",
    pages = "043193",
    year = "2022"
}

@article{Casher:1974vf,
    author = "Casher, A. and Kogut, John B. and Susskind, Leonard",
    title = "{Vacuum polarization and the absence of free quarks}",
    doi = "10.1103/PhysRevD.10.732",
    journal = "Phys. Rev. D",
    volume = "10",
    pages = "732--745",
    year = "1974"
}

@article{Banks:1976ia,
    author = "Banks, Tom and Raby, S. and Susskind, Leonard and Kogut, John B. and Jones, D. R. T. and Scharbach, P. N. and Sinclair, D. K.",
    title = "{Strong Coupling Calculations of the Hadron Spectrum of Quantum Chromodynamics}",
    doi = "10.1103/PhysRevD.15.1111",
    journal = "Phys. Rev. D",
    volume = "15",
    pages = "1111",
    year = "1977"
}

@article{Bardeen:1975gx,
    author = "Bardeen, William A. and Bars, I. and Hanson, Andrew J. and Peccei, R. D.",
    title = "{A Study of the Longitudinal Kink Modes of the String}",
    doi = "10.1103/PhysRevD.13.2364",
    journal = "Phys. Rev. D",
    volume = "13",
    pages = "2364--2382",
    year = "1976"
}

@article{Bars:1976nk,
    author = "Bars, I.",
    title = "{A Quantum String Theory of Hadrons and Its Relation to Quantum Chromodynamics in Two-Dimensions}",
    doi = "10.1016/0550-3213(76)90327-8",
    journal = "Nucl. Phys. B",
    volume = "111",
    pages = "413--440",
    year = "1976"
}

@article{tHooft:1974pnl,
    author = "'t Hooft, Gerard",
    title = "{A Two-Dimensional Model for Mesons}",
    doi = "10.1016/0550-3213(74)90088-1",
    journal = "Nucl. Phys. B",
    volume = "75",
    pages = "461--470",
    year = "1974"
}

@article{Einhorn:1976uz,
    author = "Einhorn, Martin B.",
    title = "{Form-Factors and Deep Inelastic Scattering in Two-Dimensional Quantum Chromodynamics}",
    doi = "10.1103/PhysRevD.14.3451",
    journal = "Phys. Rev. D",
    volume = "14",
    pages = "3451",
    year = "1976"
}

@article{Einhorn:1976ev,
    author = "Einhorn, Martin B.",
    title = "{Failure of the Parton Model in Inclusive electron-Positron Annihilation}",
    doi = "10.1103/PhysRevD.15.3037",
    journal = "Phys. Rev. D",
    volume = "15",
    pages = "3037",
    year = "1977"
}

@article{Shuryak:2021hng,
    author = "Shuryak, Edward and Zahed, Ismail",
    title = "{Hadronic structure on the light front. II. QCD strings, Wilson lines, and potentials}",
    doi = "10.1103/PhysRevD.107.034024",
    journal = "Phys. Rev. D",
    volume = "107",
    pages = "034024",
    year = "2023"
}

@article{Barata:2023jgd,
    author = "Barata, Jo{\~a}o and Gong, Wenjie and Venugopalan, Raju",
    title = "{Realtime dynamics of hyperon spin correlations from string fragmentation in a deformed four-flavor Schwinger model}",
    doi = "10.1103/PhysRevD.109.116003",
    journal = "Phys. Rev. D",
    volume = "109",
    pages = "116003",
    year = "2024"
}

@article{Gustafson:2024bww,
    author = "Gustafson, Erik and others",
    title = "{Surrogate Constructed Scalable Circuits ADAPT-VQE in the Schwinger model}",
    doi = "10.1103/PhysRevApplied.23.064002",
    journal = "Phys. Rev. Applied",
    volume = "23",
    pages = "064002",
    year = "2025"
}

@article{Berges:2017hne,
    author = {Berges, J\"urgen and Floerchinger, Stefan and Venugopalan, Raju},
    title = "{Dynamics of entanglement in expanding quantum fields}",
    doi = "10.1007/JHEP04(2018)145",
    journal = "JHEP",
    volume = "2018",
    number = "04",
    pages = "145",
    year = "2018"
}

@article{Florio:2021xvj,
    author = "Florio, Adrien and Kharzeev, Dmitri E.",
    title = "{Gibbs entropy from entanglement in electric quenches}",
    doi = "10.1103/PhysRevD.104.056021",
    journal = "Phys. Rev. D",
    volume = "104",
    pages = "056021",
    year = "2021"
}

@article{Grieninger:2023pyb,
    author = "Grieninger, Sebastian and Kharzeev, Dmitri E. and Zahed, Ismail",
    title = "{Entanglement entropy in a time-dependent holographic Schwinger pair creation}",
    doi = "10.1103/PhysRevD.108.126014",
    journal = "Phys. Rev. D",
    volume = "108",
    pages = "126014",
    year = "2023"
}

@article{Grieninger:2023ehb,
    author = "Grieninger, Sebastian and Kharzeev, Dmitri E. and Zahed, Ismail",
    title = "{Entanglement in a holographic Schwinger pair with confinement}",
    doi = "10.1103/PhysRevD.108.086030",
    journal = "Phys. Rev. D",
    volume = "108",
    pages = "086030",
    year = "2023"
}

@article{Berges:2017zws,
    author = {Berges, J\"urgen and Floerchinger, Stefan and Venugopalan, Raju},
    title = "{Thermal excitation spectrum from entanglement in an expanding quantum string}",
    doi = "10.1016/j.physletb.2018.01.068",
    journal = "Phys. Lett. B",
    volume = "778",
    pages = "442--446",
    year = "2018"
}

@article{Florio:2025hoc,
  title = {Thermalization from quantum entanglement: Jet simulations in the massive Schwinger model},
  author = {Florio, Adrien and Frenklakh, David and Grieninger, Sebastian and Kharzeev, Dmitri E. and Palermo, Andrea and Shi, Shuzhe},
  journal = {Phys. Rev. D},
  volume = {112},
  pages = {094502},
  year = {2025},
  doi = {10.1103/sgrx-jpp9},
}

@article{Shao:2025obi,
    author = {Shao, Haiyang and Chen, Shile and Shi, Shuzhe},
    title = {{Collective motion in the massive Schwinger model via Tensor Network}},
    journal = {arXiv:2509.10835},
    url = {http://arxiv.org/abs/2509.10835},
    year = {2025},
}

@article{Shao:2025ygy,
    author = {Shao, Haiyang and Chen, Shile and Shi, Shuzhe},
    title = {{Onset of Bjorken Flow in Quantum Evolution of the Massive Schwinger Model}},
    journal = {arXiv:2509.10855},
    url = {http://arxiv.org/abs/2509.10855},
    year = {2025},
}

@article{Grieninger:2025rdi,
  title = {Thermal nature of confining strings},
  author = {Grieninger, Sebastian and Kharzeev, Dmitri E. and Marroquin, Eliana},
  journal = {Phys. Rev. D},
  volume = {113},
  pages = {036013},
  year = {2026},
  doi = {10.1103/v8mv-185x},
}

@article{Artiaco:2025qqq,
    author = {Artiaco, Claudia and Barata, Jo{\~a}o and Rico, Enrique},
    title = {{Out-of-Equilibrium Dynamics in a U(1) Lattice Gauge Theory via Local Information Flows: Scattering and String Breaking}},
    journal = {arXiv:2510.16101},
    url = {http://arxiv.org/abs/2510.16101},
    year = {2025},
}

@article{Galvez-Viruet:2025rmy,
    author = {G{\'a}lvez-Viruet, Juan Jos{\'e} and Llanes-Estrada, Felipe J. and de Arenaza, Nicol{\'a}s Mart{\'\i}nez and G{\'o}mez-Rocha, Mar{\'\i}a and Hobbs, Timothy J.},
    title = {{Quantum Computing Hadron Fragmentation Functions in Light-Front Chromodynamics}},
    journal = {arXiv:2510.18869},
    url = {http://arxiv.org/abs/2510.18869},
    year = {2025},
}

@article{Galvez-Viruet:2025ket,
author = {G{\'a}lvez-Viruet, J. J. and Llanes-Estrada, Felipe J. and G{\'o}mez-Rocha, Mar{\'i}a},
title = {Preparations for quantum computing in hadron physics},
journal = {Int. J. Mod. Phys. A},
volume = {41},
pages = {2630003},
year = {2026},
doi = {10.1142/S0217751X26300036},
}

@article{Haegeman:2011zz,
    author = "Haegeman, Jutho and Cirac, J. Ignacio and Osborne, Tobias J. and Pizorn, Iztok and Verschelde, Henri and Verstraete, Frank",
    title = "{Time-Dependent Variational Principle for Quantum Lattices}",
    doi = "10.1103/PhysRevLett.107.070601",
    journal = "Phys. Rev. Lett.",
    volume = "107",
    pages = "070601",
    year = "2011"
}

@article{Kandala:2017vok,
    author = "Kandala, Abhinav and Mezzacapo, Antonio and Temme, Kristan and Takita, Maika and Brink, Markus and Chow, Jerry M. and Gambetta, Jay M.",
    title = "{Hardware-efficient variational quantum eigensolver for small molecules and quantum magnets}",
    doi = "10.1038/nature23879",
    journal = "Nature",
    volume = "549",
    number = "7671",
    pages = "242--246",
    year = "2017"
}

@article{Peruzzo:2014dbu,
    author = "Peruzzo, Alberto and McClean, Jarrod and Shadbolt, Peter and Yung, Man-Hong and Zhou, Xiao-Qi and Love, Peter J. and Aspuru-Guzik, Al\'an and O'Brien, Jeremy L.",
    title = "{A variational eigenvalue solver on a photonic quantum processor}",
    doi = "10.1038/ncomms5213",
    journal = "Nature Commun.",
    volume = "5",
    number = "1",
    pages = "4213",
    year = "2014"
}

@article{yastn,
      title={{YASTN: Yet another symmetric tensor networks; A Python library for Abelian symmetric tensor network calculations}},
      author={Marek M. Rams and Gabriela Wójtowicz and Aritra Sinha and Juraj Hasik},
      journal={SciPost Phys. Codebases},
      pages={52},
      year={2025},
      publisher={SciPost},
      doi={10.21468/SciPostPhysCodeb.52},
      url={https://scipost.org/10.21468/SciPostPhysCodeb.52},
}

\pagebreak

\section{Supplemental Material: {Collective organization of string fragmentation}}

This Supplemental Material collects the technical material supporting the
Letter without repeating the physical arguments developed in the main text.
We first specify the massive Schwinger model and its staggered
Kogut--Susskind Hamiltonian formulation, together with the external-charge
setup used to generate the expanding electric string
\cite{Schwinger:1962tp,Kogut:1974ag,Pichler:2015yqa,Buyens:2016ecr}.
We then give the numerical tensor-network details needed for the real-time
evolution and the analysis of midpoint electric-field probabilities and entanglement entropies of the relevant wavefunction components.
The tensor-network
and quantum-simulation literature relevant to these real-time gauge-theory
calculations is summarized in
Refs.~\cite{Martinez:2016yna,Klco:2018kyo,Zohar:2015hwa,Kasper:2016mzj,
Bauer:2022hpo,Bauer:2023qgm,Florio:2024aix,Janik:2025bbz}.

\section{The Massive Schwinger Model}

The Schwinger model is quantum electrodynamics in $1+1$-dimensions.
Although considerably simpler than QCD, it exhibits confinement,
Schwinger pair production, string breaking, and the formation of
color-singlet mesons, making it an ideal framework for studying the
real-time dynamics of hadronization.

The continuum Lagrangian is
\begin{equation}
\mathcal L=
-\frac14F_{\mu\nu}F^{\mu\nu}
+\bar\psi
\left(i\gamma^\mu D_\mu-m\right)\psi,
\qquad
D_\mu=\partial_\mu+igA_\mu ,
\end{equation}
where $g$ is the gauge coupling and $m$ is the fermion mass. In one
spatial dimension the electric field is the only independent gauge
degree of freedom. Gauss' law,
\begin{equation}
\partial_xE
=
g\,\psi^\dagger\psi+\rho_{\rm ext},
\end{equation}
relates the electric flux to the dynamical and external charge
densities. Consequently, two separated external charges are connected
by a confining electric string with continuum string tension
\begin{equation}
\kappa=\frac{g^2}{2}.
\end{equation}

The real-time evolution is computed using the staggered
(Kogut--Susskind) lattice Hamiltonian
\begin{align}
H
&=
-x
\sum_n
\left(
\chi_n^\dagger U_n\chi_{n+1}
+{\rm h.c.}
\right)
\nonumber\\
&\quad
+\mu
\sum_n
(-1)^n
\chi_n^\dagger\chi_n
+
\sum_nL_n^2 ,
\end{align}
where
\begin{equation}
x=\frac1{g^2a^2},
\qquad
\mu=\frac{2m}{g^2a},
\end{equation}
$a$ denotes the lattice spacing, $U_n=e^{i\theta_n}$ the gauge link,
and $L_n$ the conjugate electric-field operator satisfying
$[L_n,U_m]=\delta_{nm}U_n$. Physical states obey the lattice Gauss-law
constraint
\begin{equation}
L_n-L_{n-1}
=
\chi_n^\dagger\chi_n
-\frac{1-(-1)^n}{2}
+q_n^{\rm ext},
\end{equation}
where $q_n^{\rm ext}$ denotes the external charges generating the
initial electric string.

The calculations employ the matrix-product-state tensor-network
formulation as discussed in Ref.~\cite{Janik:2025bbz}. The initial state
contains an electric string stretched between two external charges that
recede along approximately lightlike trajectories. During the
subsequent real-time evolution the electric string becomes unstable
against Schwinger pair production, producing dynamical fermion pairs
that screen the electric field and reorganize into propagating
color-singlet mesons. The Letter employs the
energy-nonconserving construction introduced in
Ref.~\cite{Janik:2025bbz}, incorporating smearing of the moving external charges over neighbouring lattice sites which allows the fragmentation dynamics to be
followed over long times without introducing spurious discontinuities at $\Delta t = a$.

\section{Tensor network computation numerical details}

In all simulations, we fix $g=1$. The results presented in the plots are obtained in a finite system of $N=512$ sites for a lattice spacing $a=1/8$. The numerical convergence is checked using Matrix Product State (MPS) bond dimensions up to $1024$. We use a lattice mass $m_{\rm lat} = m - ag^2/8$~\cite{Dempsey:2022nys} which improves the convergence to the continuous limit.

Our MPS simulations employ the Hamiltonian $U(1)$ symmetry (particle number conservation), utilizing the YASTN tensor network library~\cite{yastn}. The $U(1)$ symmetry allows us to directly project onto a given value of the electric field $E$ at the bond between neighboring sites, as it is in a one-to-one correspondence with the symmetry charges spanning the virtual space of that bond. Projecting onto a given value of $E$ is equivalent to projecting a virtual MPS bond to a subspace with the corresponding value of the symmetry charge (particle number). When calculating the component entanglement entropy in Fig.~5(a) of the main text, this projected state is first normalized. As such, there is a simple relation
\begin{equation}
S_{EE} = \sum_E p_E S^E - \sum_E p_E \log_2(p_E),
\end{equation}
where $S_{EE}$ is the bipartite entanglement entropy of a given bond (shown, e.g., in Fig.~5(b) of the main text), $S^E$ is the entropy of the projected $E$ component, and $p_E$ is its probability (shown, e.g., in Fig.~3(b)).

\end{document}